\documentclass{mn2e}

\newif\ifAMStwofonts

\usepackage{graphics}
\newcommand{\ms}{\, \mathrm M_{\sun}}
\newcommand{\rs}{\, \mathrm R_{\sun}}
\newcommand{\h}{^{\mathrm h}}
\newcommand{\m}{^{\mathrm m}}
\newcommand{\s}{^{\mathrm s}}

\title{
A very massive spectroscopic binary in the LH 54 OB association in the LMC
}
\author[Pablo G. Ostrov]
{Pablo G. Ostrov\thanks{E-mail: ostrov@fcaglp.edu.ar}
\\ Facultad de Ciencias Astron\'omicas y Geof\'{\i}sicas,
Paseo del Bosque S/N (1900) La Plata, Argentina \\
}
\date{}

\pagerange{\pageref{firstpage}--\pageref{lastpage}}

\begin{document}

\maketitle

\label{firstpage}

\begin{abstract}
We announce the discovery of a new early-type,
double-lined spectroscopic binary in the LH 54 OB association in the LMC.
We present a $V$ light curve and radial velocities.
We investigate the possible configurations of the system, concluding that it
probably contains the most massive star measured at the present, with a mass 
of the order of $100 \ms$, while its companion has approximately $50 \ms$.
\end{abstract}

\begin{keywords}
binaries: eclipsing -- stars: early-type -- stars: fundamental 
parameters -- stars: individual: LH 54-425
\end{keywords}

\section{Introduction}

Though we can trust that the present development of stellar interiors 
modelling reproduces qualitatively the consecutive stages along the evolution 
of stars, there  still remain appreciable quantitative discrepancies between 
theoretical predictions and empirical evidence, especiallly in that 
concern to the higher mass stars.
That is not surprising since, on one hand, their modelling is intrinsically 
more difficult due to their more complex evolution, and on the other hand,
very massive stars are rare because of its shorter lifetimes, so that
there are scarce observational constraints.

During the last years, we have performed a campaign of observations of early 
Magellanic stars in order to enrich the empirical knowledge of these objects,
particularly in low metallicity regions.

One of the new binaries discovered for which its light curve was completed
is a $V \sim 13$ mag star in the LH 54 OB association (Lucke \& Hodge 1970).
This star ($\alpha=5\h26\m24\s$, $\delta=-67\degr30\arcmin13\arcsec$)
is bright enough and has sufficiently big radial velocity 
excursions to be at scope of the 2.15-m telescope at CASLEO\footnote{
Complejo Astron\'omico El Leoncito, operated under agreement between 
the Consejo Nacional de Investigaciones Cient\'{\i}ficas y T\'ecnicas de la 
Rep\'ublica Argentina and the National Universities of La Plata, C\'ordoba and 
San Juan.}
(San Juan, Argentina) for radial velocity measurements.

Hill et al. (1994) identified this object as LH 54-425 and performed $UBV$
photometry, obtaining $V=13.19 \pm 0.01$, $B-V=-0.31 \pm 0.01$ and 
$U-B=-1.02 \pm 0.01$.  Later on, Oey (1996a) obtained 
$V=13.13 \pm 0.007$, $B-V=-0.215 \pm 0.014$ and $U-B=-1.010 \pm 0.015$,
listing this star as L54S-4.
CCD spectra were also obtained by 
Oey (1996b) with the ARGUS multifiber spectrograph at the 
CTIO 4-m telescope. The spectral type assigned by Oey to this star was O4 III 
(f$^{\ast}$). She did not detect double lines at the phase of her observation.

LH 54-425 is the earliest star of the OB association LH 54, located at the 
East-side of the superbubble DEM 192 (Davies, Elliot 
\& Meaburn 1976) (or N51D, Henize 1956) that is probably related with 
its growth (Oey \& Smedley 1998).

\section{Observations}

LH 54 had been included in one of our selected fields for searching new 
eclipsing binaries. It was observed during four observing runs between 1998 
and 2001.

LH 54-425 showed no eclipses, but exhibited small periodic luminosity changes 
of the order of $\sim 0.1$ mag, suggesting an ellipsoidal nature of the 
variability. For this reason we included it as subject for further 
spectroscopic study as soon as its light curve was sufficiently complete to 
compute a reasonably reliable ephemeris. 






\subsection{Photometry}

CCD images of the region of LH 54 were obtained at CASLEO between 1998 and 
2001, with the same instrumentation described in Ostrov et al. (2000).
We derived a $V$ light curve from aperture photometry performed with DAOPHOT
(Stetson 1987, 1991), and tied all the measurements to a unique 
instrumental system using several stars as local standards to reduce errors, 
as described in Ostrov et al. (2000). 
We performed absolute photometry during a photometric night (Dec 3, 1998), 
obtaining $V=13.05 \pm 0.015$, and $B-V=-0.121 \pm 0.011$.
Table \ref{foto} displays our photometric measurements of LH 54-425, 
corrected to the standard system, together with the internal errors derived
from the local standards, the $FWHM$ and the airmass for each frame.

\begin{table*}  
\caption[]{$V$ light photometry of LH 54-425}
\label{foto}
  
\begin{tabular}{rrrrlrrrrl}  
\hline  
\noalign{\smallskip}
  
 HJD       &$    V    $&$ \sigma_{\rm i}$&   fwhm    &$  X    $&
 HJD       &$    V    $&$\sigma_{\rm i}$&   fwhm    &$  X    $\cr  
$  2450000+ $&           &               &$  \arcsec  $&               &
$  2450000+ $&           &               &$  \arcsec  $&               \cr
\noalign{\smallskip}  
\hline  
\noalign{\smallskip}  
$  1149.807 $&$  13.087 $&$   0.011 $&$   2.09 $&$   1.30 $&$  1503.601 $&$  13.048 $&$   0.008 $&$   2.62 $&$   1.48 $ \cr
$  1149.834 $&$  13.061 $&$   0.011 $&$   2.36 $&$   1.36 $&$  1503.687 $&$  13.058 $&$   0.006 $&$   2.86 $&$   1.27 $ \cr
$  1150.847 $&$  13.082 $&$   0.012 $&$   1.91 $&$   1.40 $&$  1503.742 $&$  13.059 $&$   0.007 $&$   2.40 $&$   1.23 $ \cr
$  1151.617 $&$  13.067 $&$   0.006 $&$   1.92 $&$   1.33 $&$  1503.795 $&$  13.069 $&$   0.005 $&$   2.84 $&$   1.25 $ \cr
$  1151.671 $&$  13.059 $&$   0.008 $&$   2.04 $&$   1.25 $&$  1503.836 $&$  13.069 $&$   0.004 $&$   2.83 $&$   1.30 $ \cr
$  1151.731 $&$  13.050 $&$   0.004 $&$   2.00 $&$   1.23 $&$  1504.575 $&$  13.048 $&$   0.006 $&$   3.46 $&$   1.58 $ \cr
$  1151.786 $&$  13.056 $&$   0.006 $&$   2.13 $&$   1.28 $&$  1504.628 $&$  13.046 $&$   0.007 $&$   3.14 $&$   1.39 $ \cr
$  1151.857 $&$  13.048 $&$   0.006 $&$   2.05 $&$   1.44 $&$  1504.736 $&$  13.050 $&$   0.008 $&$   3.49 $&$   1.23 $ \cr
$  1152.663 $&$  13.085 $&$   0.005 $&$   2.26 $&$   1.26 $&$  1504.776 $&$  13.051 $&$   0.007 $&$   2.72 $&$   1.24 $ \cr
$  1152.709 $&$  13.076 $&$   0.007 $&$   1.98 $&$   1.23 $&$  1504.814 $&$  13.053 $&$   0.006 $&$   3.00 $&$   1.27 $ \cr
$  1152.751 $&$  13.073 $&$   0.005 $&$   2.19 $&$   1.24 $&$  1504.842 $&$  13.060 $&$   0.007 $&$   3.16 $&$   1.32 $ \cr
$  1152.800 $&$  13.066 $&$   0.006 $&$   1.98 $&$   1.31 $&$  1859.592 $&$  13.058 $&$   0.007 $&$   3.62 $&$   1.63 $ \cr
$  1152.831 $&$  13.062 $&$   0.006 $&$   1.86 $&$   1.37 $&$  1859.615 $&$  13.059 $&$   0.005 $&$   3.18 $&$   1.53 $ \cr
$  1153.692 $&$  13.105 $&$   0.007 $&$   1.62 $&$   1.23 $&$  1859.652 $&$  13.052 $&$   0.006 $&$   3.04 $&$   1.40 $ \cr
$  1153.735 $&$  13.106 $&$   0.006 $&$   1.76 $&$   1.24 $&$  1859.683 $&$  13.052 $&$   0.004 $&$   3.51 $&$   1.32 $ \cr
$  1153.768 $&$  13.092 $&$   0.006 $&$   1.87 $&$   1.26 $&$  1859.709 $&$  13.054 $&$   0.005 $&$   4.02 $&$   1.28 $ \cr
$  1153.823 $&$  13.079 $&$   0.005 $&$   2.36 $&$   1.36 $&$  1859.721 $&$  13.045 $&$   0.019 $&$   6.23 $&$   1.26 $ \cr
$  1153.857 $&$  13.070 $&$   0.008 $&$   2.40 $&$   1.46 $&$  1859.763 $&$  13.063 $&$   0.013 $&$   6.23 $&$   1.23 $ \cr
$  1154.706 $&$  13.110 $&$   0.005 $&$   2.06 $&$   1.23 $&$  1859.806 $&$  13.046 $&$   0.004 $&$   5.11 $&$   1.24 $ \cr
$  1154.740 $&$  13.100 $&$   0.005 $&$   2.16 $&$   1.24 $&$  1859.838 $&$  13.047 $&$   0.007 $&$   4.63 $&$   1.26 $ \cr
$  1154.804 $&$  13.102 $&$   0.007 $&$   2.48 $&$   1.32 $&$  1860.592 $&$  13.092 $&$   0.010 $&$   5.08 $&$   1.62 $ \cr
$  1154.845 $&$  13.096 $&$   0.004 $&$   2.45 $&$   1.43 $&$  1860.687 $&$  13.073 $&$   0.006 $&$   5.37 $&$   1.31 $ \cr
$  1155.674 $&$  13.105 $&$   0.005 $&$   2.03 $&$   1.24 $&$  1860.760 $&$  13.065 $&$   0.005 $&$   3.87 $&$   1.23 $ \cr
$  1155.712 $&$  13.111 $&$   0.004 $&$   1.99 $&$   1.23 $&$  1860.832 $&$  13.070 $&$   0.006 $&$   5.22 $&$   1.26 $ \cr
$  1155.748 $&$  13.115 $&$   0.005 $&$   2.05 $&$   1.25 $&$  1860.867 $&$  13.053 $&$   0.005 $&$   4.74 $&$   1.31 $ \cr
$  1155.776 $&$  13.116 $&$   0.005 $&$   1.86 $&$   1.28 $&$  1861.604 $&$  13.109 $&$   0.007 $&$   3.65 $&$   1.55 $ \cr
$  1155.826 $&$  13.115 $&$   0.007 $&$   2.49 $&$   1.38 $&$  1861.733 $&$  13.092 $&$   0.005 $&$   4.05 $&$   1.25 $ \cr
$  1155.851 $&$  13.108 $&$   0.006 $&$   2.56 $&$   1.45 $&$  1861.815 $&$  13.075 $&$   0.004 $&$   3.37 $&$   1.25 $ \cr
$  1499.825 $&$  13.090 $&$   0.004 $&$   2.90 $&$   1.27 $&$  1861.860 $&$  13.067 $&$   0.006 $&$   3.66 $&$   1.31 $ \cr
$  1500.589 $&$  13.089 $&$   0.011 $&$   3.61 $&$   1.56 $&$  1862.572 $&$  13.112 $&$   0.007 $&$   3.65 $&$   1.69 $ \cr
$  1500.659 $&$  13.101 $&$   0.004 $&$   2.89 $&$   1.34 $&$  1862.638 $&$  13.109 $&$   0.006 $&$   3.44 $&$   1.42 $ \cr
$  1500.732 $&$  13.107 $&$   0.004 $&$   3.01 $&$   1.24 $&$  1862.698 $&$  13.109 $&$   0.006 $&$   3.26 $&$   1.28 $ \cr
$  1500.799 $&$  13.110 $&$   0.007 $&$   3.19 $&$   1.24 $&$  1862.751 $&$  13.099 $&$   0.005 $&$   3.36 $&$   1.23 $ \cr
$  1500.844 $&$  13.104 $&$   0.006 $&$   3.34 $&$   1.30 $&$  1862.833 $&$  13.092 $&$   0.005 $&$   3.53 $&$   1.27 $ \cr
$  1501.549 $&$  13.068 $&$   0.005 $&$   3.30 $&$   1.76 $&$  1863.564 $&$  13.104 $&$   0.005 $&$   2.64 $&$   1.72 $ \cr
$  1501.593 $&$  13.077 $&$   0.007 $&$   3.38 $&$   1.53 $&$  1863.634 $&$  13.117 $&$   0.007 $&$   3.08 $&$   1.42 $ \cr
$  1501.616 $&$  13.079 $&$   0.007 $&$   3.25 $&$   1.45 $&$  1863.720 $&$  13.118 $&$   0.004 $&$   2.69 $&$   1.25 $ \cr
$  1501.677 $&$  13.090 $&$   0.004 $&$   3.10 $&$   1.30 $&$  1863.810 $&$  13.116 $&$   0.004 $&$   3.45 $&$   1.25 $ \cr
$  1501.756 $&$  13.098 $&$   0.005 $&$   2.60 $&$   1.23 $&$  1864.619 $&$  13.084 $&$   0.006 $&$   2.27 $&$   1.46 $ \cr
$  1501.815 $&$  13.106 $&$   0.004 $&$   3.00 $&$   1.26 $&$  1864.681 $&$  13.093 $&$   0.006 $&$   1.79 $&$   1.30 $ \cr
$  1502.565 $&$  13.047 $&$   0.008 $&$   3.93 $&$   1.65 $&$  1864.755 $&$  13.108 $&$   0.005 $&$   2.38 $&$   1.23 $ \cr
$  1502.699 $&$  13.078 $&$   0.009 $&$   2.36 $&$   1.26 $&$  1864.812 $&$  13.119 $&$   0.006 $&$   2.80 $&$   1.25 $ \cr
$  1502.784 $&$  13.083 $&$   0.008 $&$   2.12 $&$   1.24 $&$  2212.865 $&$  13.071 $&$   0.006 $&$   2.82 $&$   1.26 $ \cr
$  1502.832 $&$  13.089 $&$   0.008 $&$   2.14 $&$   1.29 $&$  2213.797 $&$  13.060 $&$   0.005 $&$   2.74 $&$   1.23 $ \cr
$  1502.861 $&$  13.096 $&$   0.018 $&$   3.99 $&$   1.34 $&$  2214.855 $&$  13.062 $&$   0.006 $&$   1.99 $&$   1.25 $ \cr
\noalign{\smallskip}  
\hline  
\end{tabular}
\end{table*}  

\subsection{Spectroscopy}

Spectroscopic observations were obtained by means of a REOSC spectrograph 
in its single dispersion mode, attached to the 2.15-m telescope at CASLEO.
The CCD detector was the same Tektronik used for photometry. We used a 600
mm$^{-1}$ grating giving a reciprocal dispersion of 1.63 \AA/pixel on the 
range from 3900 to 5500 \AA. 
Our resolution, measured from the Cu-Ne-Ar
comparison lamp lines, was 2 pixels.

We obtained spectra near the quadratures in series of three 1200 s exposures.
The spectrograms were processed and extracted using IRAF\footnote{
IRAF software is distributed by NOAO, operated by AURA for NSF.}
routines.
Each series of three observations was combined in one spectrogram, before
performing the concomitant measurements.



\section{Analysis}

\subsection{Ephemeris}

We applied to the photometric data two period search methods: one derived from 
the Lafler \& Kinman's (Lafler \& Kinman 1965) and the one described in 
Shwarzenberg-Czerny (1997). Because of our short time base and the 
intrinsically small light variations, we can not derive a precise ephemeris. 
We found that the most probable period is $P=2.2475$ days, with aliases each 
0.007 days. We used this period and $E_0=2452216.597$ to plan the
spectroscopic observations.

\subsection{Spectroscopic classification}

\begin{figure*}
  \resizebox{12cm}{!}{\includegraphics{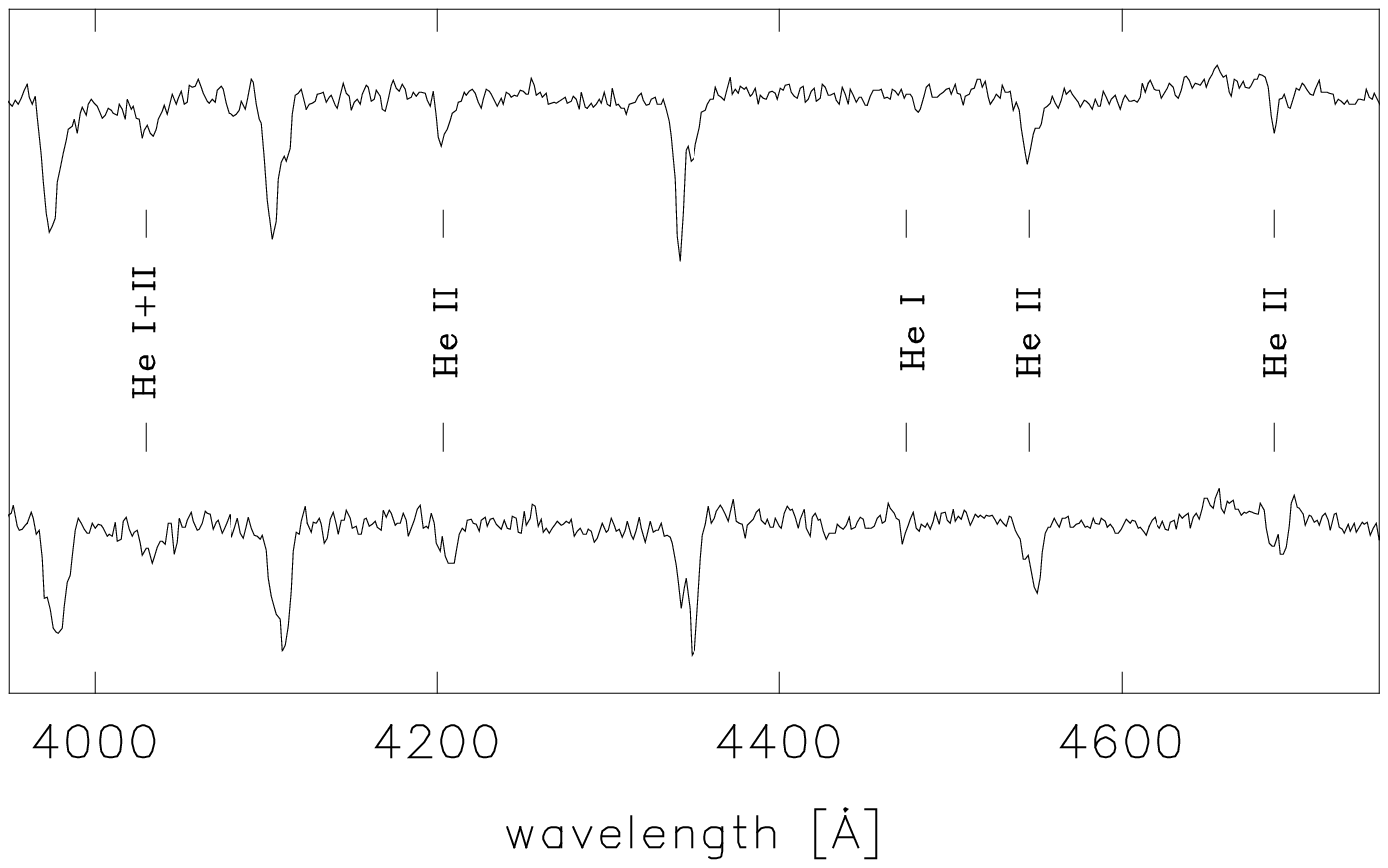}}
  \caption[]{
CCD spectra of LH 54-425 obtained at CASLEO, corresponding to both quadratures.
}
 \label{spectra}
\end{figure*}

We classified the spectra following the criteria of  Walborn \& Fitzpatrick 
(1990). The earliest component shows no lines of He {\sc i}, 
corresponding thus an O3 spectral type. The other component shows He {\sc i} 
4471 hardly visible due our poor signal to noise ratio (in fact, it is only 
detectable in our best spectrograms, obtained during dark nights) and more 
intense He~{\sc ii} lines which is in agreement with an approximately O5 type. 
With regard to  the luminosity 
classes, we combined all the spectrograms acquired in each quadrature to 
increment the signal to noise ratio. In the averaged spectrograms, 
N {\sc iv} 4058 emission
from the primary is visible, together with Si {\sc iv} $\lambda \lambda$ 
4089 or  4116 depending on the phase (this is alternatively blended with the 
H$\delta$ absorption of the secondary). We are not able to assign a luminosity
class to the secondary. So, we classify the system as O3 III(f$^{\ast}$)+O5:.
Fig. \ref{spectra} shows the CCD spectrograms of LH 54-425 obtained
during dark nights, corresponding to both quadratures.

\subsection{Radial velocity measurements}

The He {\sc ii} 4686 line appears 
clearly double 
in all our spectrograms.
He {\sc ii} 4200 and He {\sc ii} 4542 appear double in some spectrograms and 
severely 
blended in others.
The He {\sc ii} absorption pairs were measured by means
of a double gaussian fit using the ``deblend'' function of the ``splot'' IRAF
task.
Only the easily separable He {\sc ii} pairs were used for the radial 
velocity determinations.
Table \ref{vr} lists the heliocentric radial velocities 
derived for both binary components, together with that corresponding to
the nebular [O {\sc iii}]$\lambda$ 5007 emission, used to check the stability 
of the system. 

\begin{table}
\label{vr}
 \centering
  \caption{Radial velocities}
  \begin{tabular}{@{}lcccc@{}}
\hline
HJD-2450000 & phase & Primary & Secondary & [O{\sc iii}] \cr
&& [km s$^{-1}$] & [km s$^{-1}$] & [km s$^{-1}$] \cr
\hline
2286.732 & 0.702 & 122	& ~679  & 286 \cr
2287.721 & 0.146 & 500  & ~~19  & 286 \cr
2329.516 & 0.742 & 114	& ~704  & 269 \cr
2330.537 & 0.197 & 496	& -169  & 280 \cr
2330.645 & 0.245 & 450	& -119  & 280 \cr
2331.546 & 0.646 & ~96	& ~648  & 284 \cr
\hline
\end{tabular}
\end{table}

\subsection{Geometry system's exploration}

Since the system shows no true eclipses, it is not possible to derive 
reliable physical parameters just from the light curve and radial velocities.
For this reason, we assumed an absolute magnitude
and investigate which system configurations were compatible with the 
observations.

We computed the models with the Wilson-Devinney codes (Wilson 1990, Wilson \& 
Devinney 1971). For the bolometric albedos and gravity darkening coefficients 
we assumed values of $A=1.0$ and $g=1.0$, respectively, which are adequate for 
radiative envelopes (Rucinski 1969, Lucy 1976). For the limb darkening a 
square root law (D\'{\i}az-Cordov\'es \&  Gim\'enez, 1992) was used,
taking the corresponding coefficients from tables by D\'{\i}az-Cordov\'es 
et al. (1995).

\subsubsection{Detached configurations}

We first explored the possible detached configurations, so the mode of 
operation 2 of the Wilson-Devinney program was chosen. In order to reduce the
degrees of freedom of the problem, we fixed the total bolometric magnitude of 
the system. We considered three cases, with $M_{bol}=-10, -10.3$ and $-10.6$.
For the models with $M_{bol}=-10$ we assigned temperatures of 52500 K and 
44500 K, corresponding to spectral types O3\,V and O5\,V according with 
Schmidt-Kaler (1982), while for those models with $M_{bol}=-10.6$ we adopted 
the temperature scale of Chlebowski \& Garmany (1991) for O3\,III and O5\,III 
stars, 46500 K and 42300, respectively. For the models with intermediate
bolometric magnitude we used the mean of the two temperature scales.
With this procedure, the extreme cases in distance moduli and temperature
scales were considered.

For each adopted orbital inclination, temperatures and potential $\Omega_1$ of 
the primary, the secondary's potential $\Omega_2$ was chosen according with
the assumed total bolometric magnitude of the system.

\begin{figure}
  \resizebox{7cm}{!}{\includegraphics{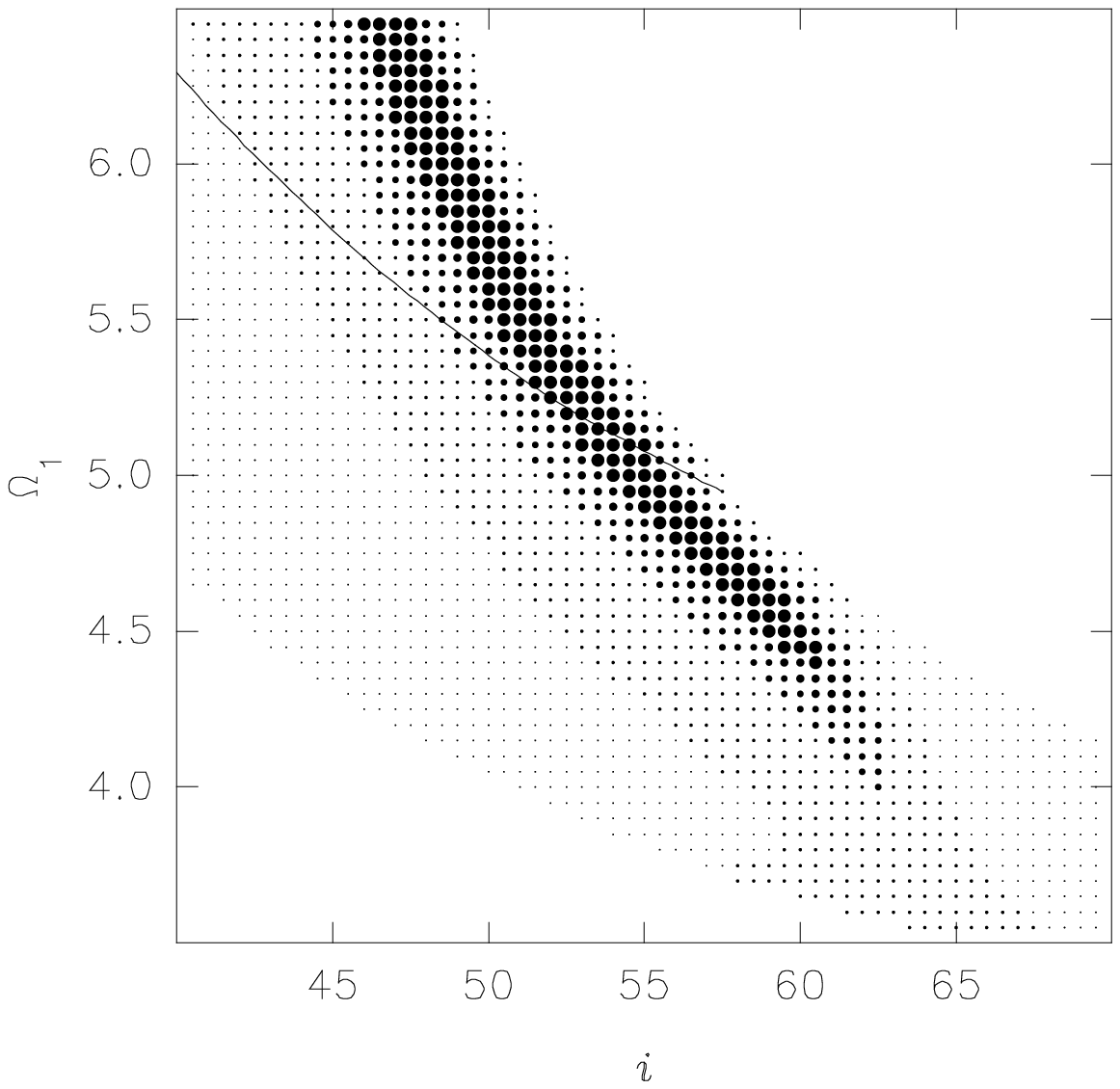}}
  \resizebox{7cm}{!}{\includegraphics{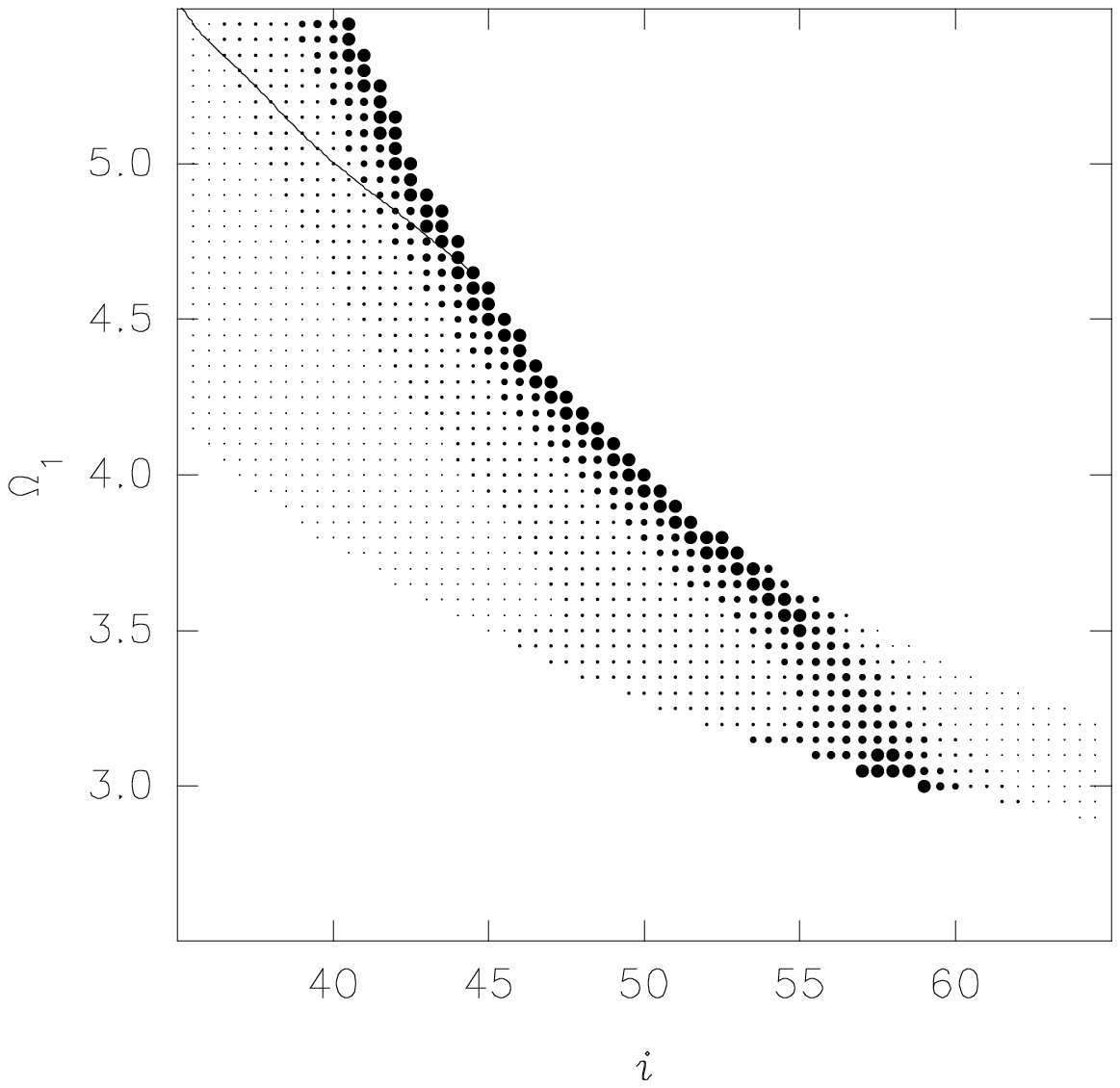}}
  \resizebox{7cm}{!}{\includegraphics{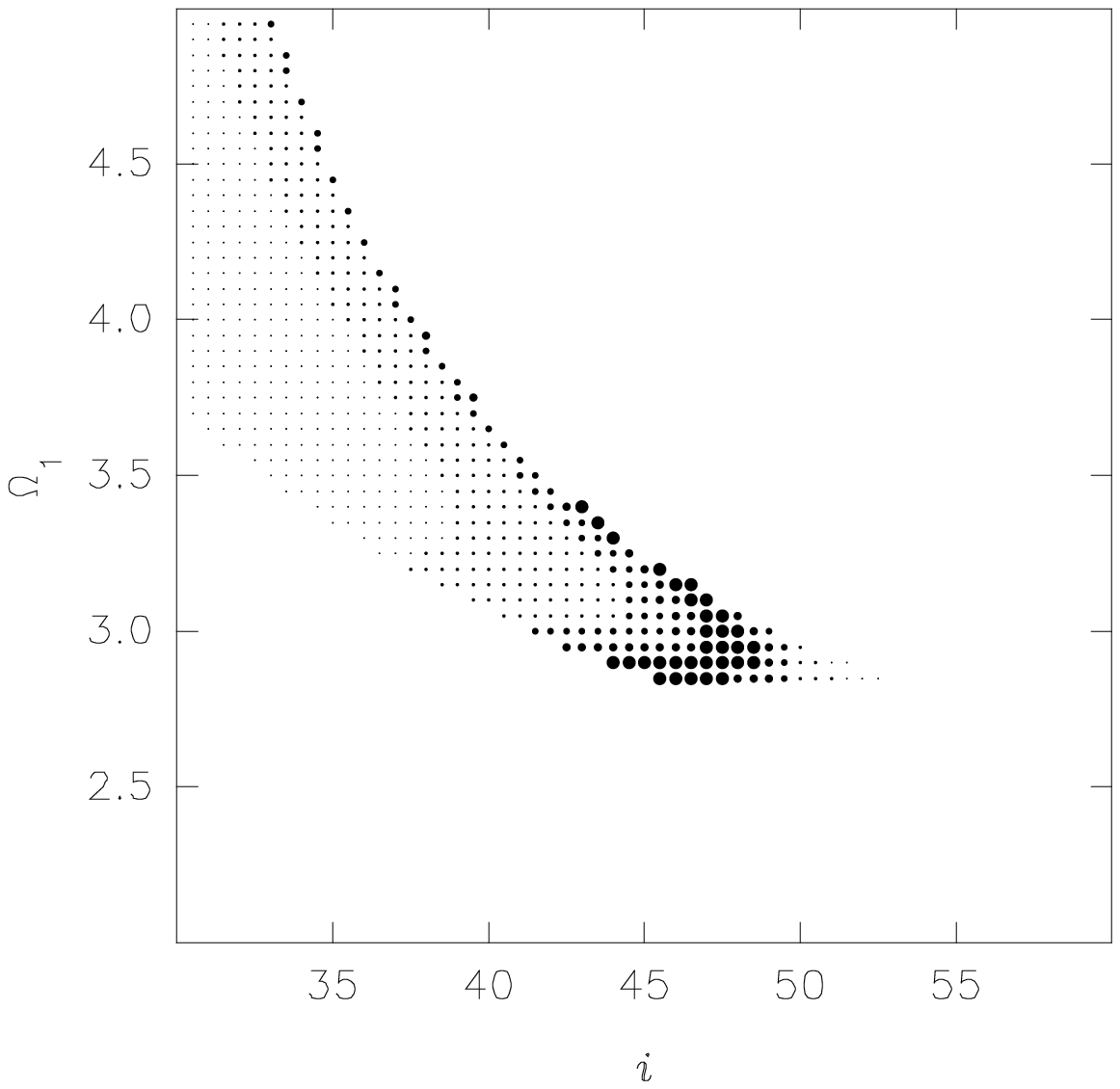}}
  \caption[]{Residuals of the detached models. Points of different sizes
represent the sums of squared residuals, the bigger symbols corresponding to
the smaller amounts. The limits between different point sizes are 1.65, 3.30, 
4.95 and 8.24 times the sum of squared residuals of the best fitting model.
%
The different panels correspond to different assumed effective temperatures and
total luminosity of the system, as quoted in the text.
}
  \label{det}
\end{figure}

Fig. \ref{det} displays the results of the experiments with 
detached configurations. Points of bigger size represent smaller O-C
light curve residuals.
The lower limit corresponds to the primary being 
2.5 magnitudes brighter than the secondary component,  which can be rejected
according to the relative intensity of the corresponding spectral lines.
The upper limit
corresponds to semi-detached configurations, with the secondary star filling
its Roche-lobe. Models above the solid curve have the secondary brighter than
the primary, therefore can be also disregarded. 

The upper panel, that corresponds to highest temperatures and lower luminosity,
shows that the best fitting models are near the semi-detached configuration,
with the secondary filling its Roche-lobe. 
These models are not realistic from
an evolutionary perspective, since they have secondary components of bigger 
size than the primary. If the less massive star is the more evolved one,
we have to conclude that mass inversion has occurred, and the system is now
semi-detached.

In the middle panel we considered models with intermediate luminosities and
temperatures. The best solutions are aligned at the upper limit of the models
considered, indicating a semi-detached configuration. There are also good 
models with $i \sim 58 \deg$ and $\Omega_1 \sim 3$. It means that the 
{\it primary} star is almost filling its Roche-lobe. However, it is not very 
plausible since the magnitude difference between both stars is near 2.5 mag, 
which does not agree with the spectroscopic evidence
\footnote{
It does not allows to disregard completely such solutions. For example, 
Niemel\"a \& Bassino (1994) called ``primary'' at the Roche-lobe filling
component of HV\,2543, based on the intensity of their spectral lines,
while from the W-D modelling that star is approximately 0.9 $V$ magnitudes 
fainter than its companion.}.

The residuals from models with lowest temperature scale and highest total 
luminosity (which allows the biggest sizes of the stars) are presented in the
bottom panel. In this case, besides of the semi-detached configurations, there
exist plausible detached solutions, with $i \sim 47 \deg$. The lowest value
considered for the potential $\Omega_1$ (2.85) implies that the primary 
component is almost filling its Roche-lobe, although (as mentioned above) in 
these models the luminosity difference between both stars is excessive.

%

\subsubsection{Semi-detached configurations}

\begin{figure}
  \resizebox{7cm}{!}{\includegraphics{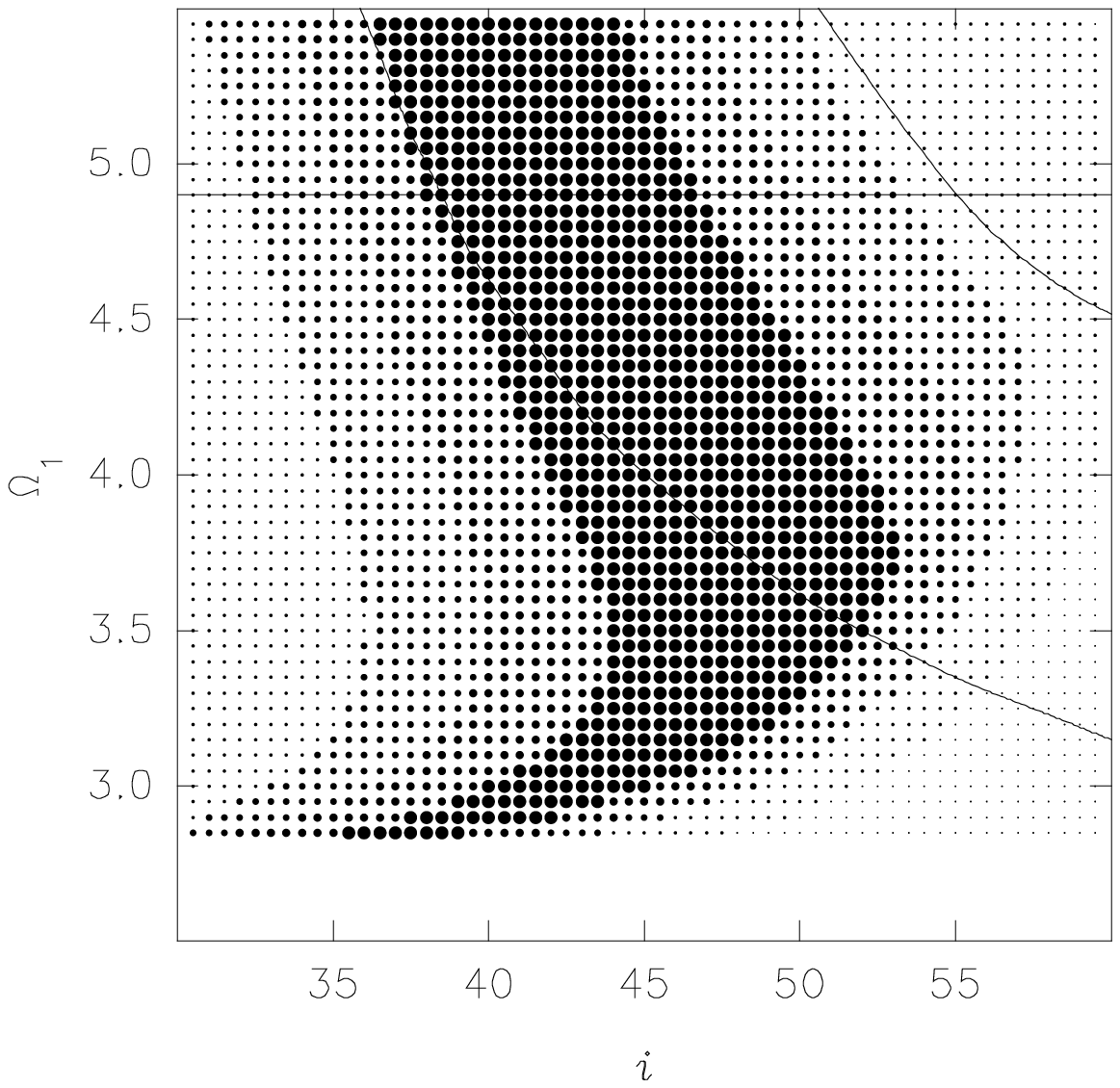}}
  \resizebox{7cm}{!}{\includegraphics{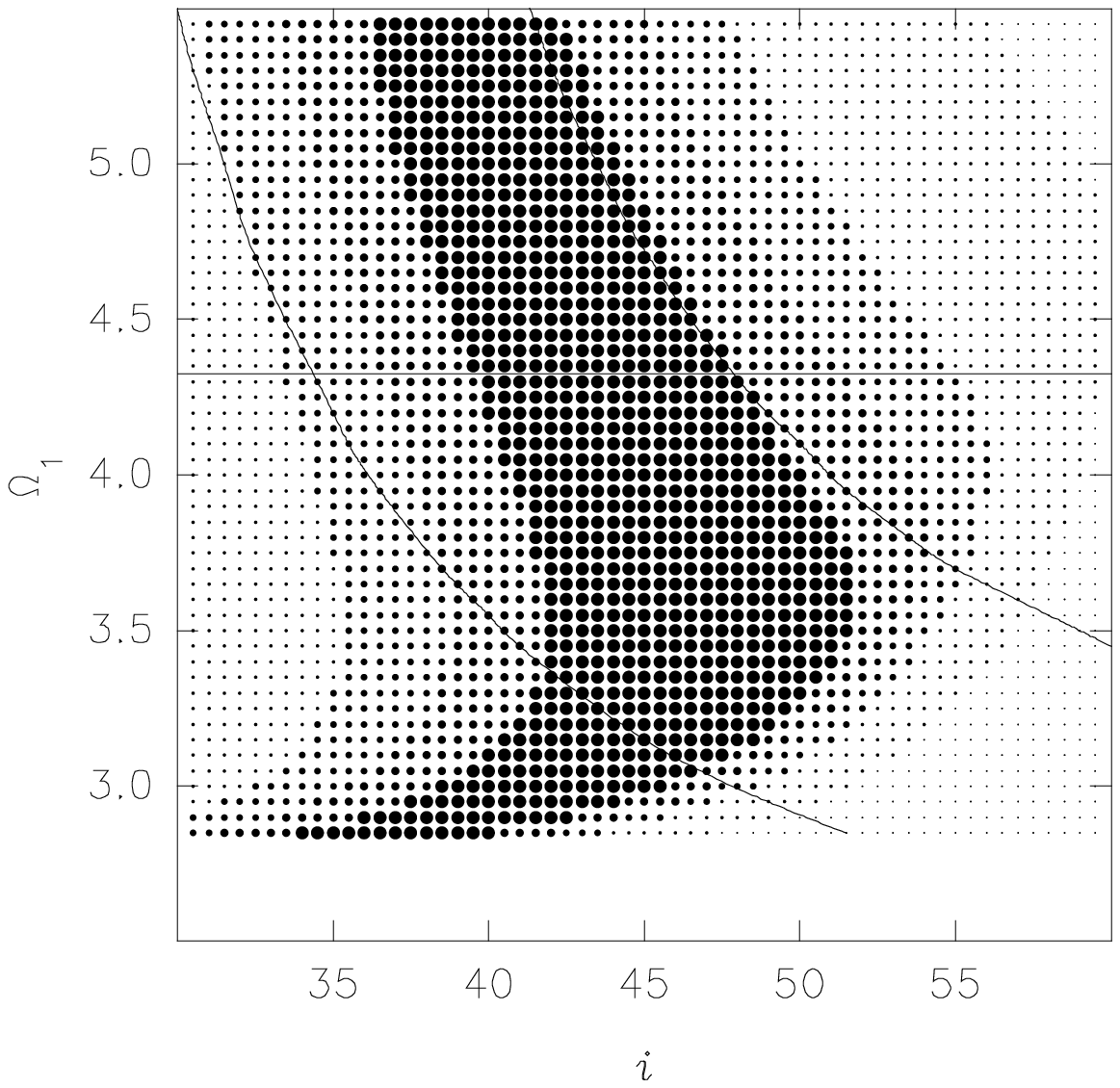}}
  \caption[]{Residuals of the semi-detached models. Points of different sizes
represent the sums of squared residuals, as in Fig. \ref{det}.
Top and bottom panels correspond to different assumed temperature scales, 
as quoted in the text.
}
	\label{sd}
\end{figure}

Considering that some results of the experiments suggested that the system
is semi-detached, with the cooler and less massive star filling its Roche-lobe,
we explored these configurations, using therefore the mode of operation 5 in
the Wilson-Devinney programs. 
Figure \ref{sd} shows the residuals of these models,
corresponding to the high and low temperature scales above mentioned, 
respectively. As the margins of plausible models are wide, it is necessary to 
consider limits to the total luminosity to restrict the amount of solutions.
The upper and bottom curves drawn on figure \ref{sd} correspond
to models with $M_{bol}=-10$ and $M_{bol}=-10.6$, respectively, while models
above the horizontal line have the secondary more luminous than the primary
component. Configurations with $\Omega_1$ smaller than those plotted in the
graphics are in overcontact, and can be also rejected according to the 
assumed limits in the total luminosity of the system.

\section{Results}

In spite of the wide range of models that can fit well the observations, we can
state with reasonable confidence that the inclination is roughly 
$45 \sim 50 \degr$, 
since bigger values would cause eclipses.
This implies that the masses are 
$\sim 100$ and $\sim 50 M_{\sun}$ for the O3 and the O5 component, 
respectively. Table \ref{resum} summarises the parameters that best fit 
suitable detached and semi-detached solutions. 
It must be noted that the errors
quoted in Table \ref{resum} correspond to fixed values of $i$, $\Omega_1$
and $\Omega_2$, i.e., the errors of the masses reflect only the errors
in the radial velocity measurements.

\begin{figure}
  \resizebox{\hsize}{!}{\includegraphics{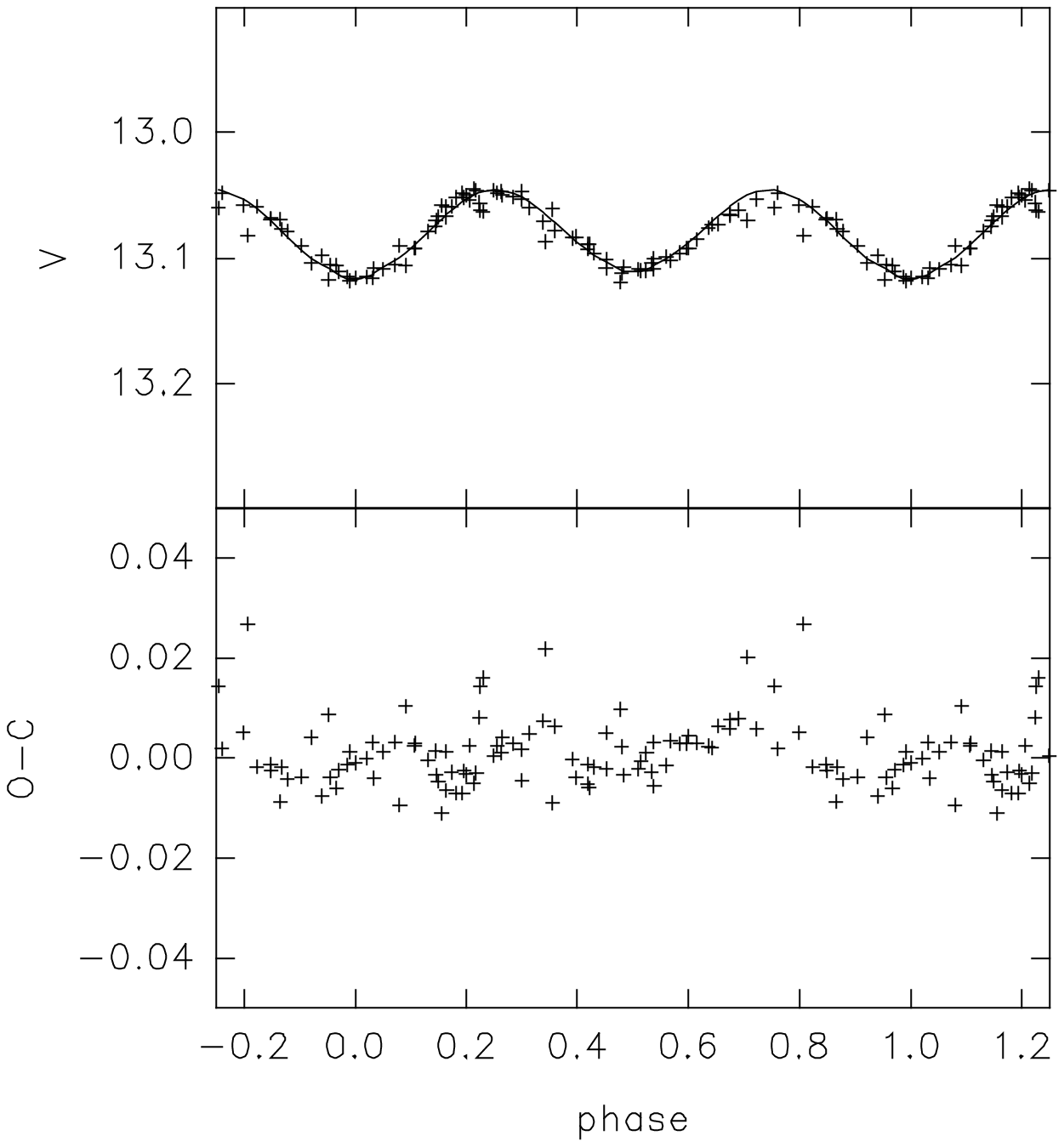}}
  \caption[]{The observed and modelled $V$ light curve for LH 54-425 and their
concomitant O-C residuals.
}
	\label{lc}
\end{figure}

\begin{figure}
  \resizebox{\hsize}{!}{\includegraphics{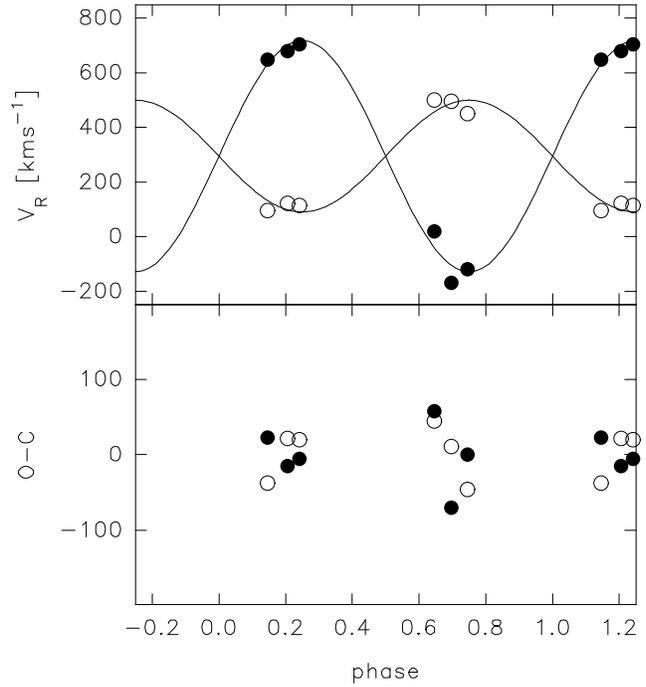}}
  \caption[]{Observed and modelled radial velocity curve for LH 54-425 and
their O-C residuals.
Hollow circles correspond to the primary component and filled ones stand for 
the secondary.
}
	\label{vc}
\end{figure}

The modelled light and radial
velocity curves are displayed in Fig. \ref{lc} and Fig. \ref{vc}, together
with the observations and O-C residuals. The models correspond to the best 
fitting detached configuration, although there are not visible difference
with the best fitting semi-detached solution.

\begin{table}
\label{resum}
 \centering
  \caption{Best fitting plausible models}
  \begin{tabular}{@{}lrr@{}}
\hline
\multicolumn{3}{c}{Adopted Values} \cr
\noalign{\medskip} 
$T_1$                  &&$ 46500\,{\rm K}$ \cr 
$T_2$                  &&$ 42300 \, {\rm K}$  \cr    
$M_{bol}$	       &&$ -10.6$ \cr 
\noalign{\medskip} 
\multicolumn{3}{c}{Fitted Values} \cr
\noalign{\medskip} 
$V_{\gamma}$           &&$ 295.0 \pm 3 \, {\rm km~s}^{-1}    $ \cr  
$q~(M_2/M_1)$          &&$ 0.48 \pm 0.01    $ \cr 
\hline
configuration & detached & semi-detached  \cr
\noalign{\medskip} 
$i$ (adopted)		&$ 47 \degr$ &$ 45.5 \degr$\cr
$\Omega_1$ (adopted)	&$ 3.1 $ &$ 3.15$  \cr 
$\Omega_2$   		&$ 2.86^{\rm a}$ &$ 2.84^{\rm b}$\cr
$a$                    &$ 38.1 \pm 0.4 \rs$ &$ 39.3 \pm 0.4 \rs  $\cr 
$\Sigma res^2$	& $0.0033222$ & $0.0033563$ \cr
\noalign{\medskip} 
\multicolumn{3}{c}{Star Dimensions} \cr
\noalign{\medskip} 
$M_1       $&$  100 \pm 3~\ms  $&$  108 \pm 3~\ms  $\cr
$R_1       $&$  15.1 \rs $&$  15.1 \rs  $\cr  
$M_{\rm bol~1} $&$ -10.2     $&$ -10.2     $\cr        
$\log g_1 [cgs]  $&$ 4.1 \pm 0.1$&$ 4.1 \pm 0.1$\cr  
\noalign{\smallskip}
$M_2   $&$  48.4 \pm 3~ \ms  $&$  52.2 \pm 3~ \ms  $\cr
$ R_2  $&$  11.9 \rs   $&$  12.4 \rs   $\cr  
$ M_{\rm bol~2} $&$ -9.3   $&$ -9.3   $\cr        
$\log g_2 [cgs]   $&$ 4.0 \pm 0.1   $&$ 4.0 \pm 0.1   $\cr  
\hline
\end{tabular}
      \begin{list}{}{}  
      \item[$^{\rm a}$] Chosen according the total luminosity of the system. 
 	\item[$^{\rm b}$] $\Omega_{\rm crit}$
    \end{list}    
\end{table}

In the more 
modest case, the inclination is $i \sim 58 \degr$ and the corresponding masses
are $\sim 64$ and $\sim 31 M_{\sun}$, although this model have a primary 2.3
magnitudes brighter than the secondary component, that is not supported by
the spectroscopic data. 
(The other extreme case, in that both stars have 
similar brightness, corresponds 
to $i \sim 39 \degr$ and gives masses of $\sim 160$ and $\sim 80 M_{\sun}$).
All possible solutions
indicate that this system contains the more massive star measured at the
present.

In a previous work (Ostrov 2001) we found from the analysis of another 
early Magellanic 
binary with O3If$^{\ast}$ and O6V components, masses of 41 and 27 $M_{\sun}$.
That was an overcontact system, with a still shorter period, of only 1.4 
days. These results suggests that extremely early stars have a wide range of 
masses, at least when they belong to very close binaries, where mass exchange
and mass loss play a fundamental role in the evolution.
 
Indeed, observations of these systems with higher signal to noise and 
dispersion than that  reachable at CASLEO would provide valuable data to 
investigate the similitudes and differences between these objects.


\section*{Acknowledgments}

I wish especially to thank Dr. Claudia Giordano for help with English,
and Dr. Nidia Morrell for helpful comments.
The author acknowledge use at CASLEO of the CCD and data acquisition system   
supported through U.S. National Science Foundation grant AST-90-15827 to   
R. M. Rich. The focal reducer in use at CASLEO was kindly provided by Dr. M.
Shara. This research has made use of the Astronomical Data Center 
catalogs. 
I would also like to thank the referee, Dr Otmar Stahl,
for suggestions that allowed to 
clarify some issues in the paper.

\bsp

\label{lastpage}

\end{document}